\documentclass{article}

\usepackage{arxiv}

\usepackage[utf8]{inputenc} 
\usepackage[T1]{fontenc}    
\usepackage{hyperref}       
\usepackage{url}            
\usepackage{booktabs}       
\usepackage{amsfonts}       
\usepackage{amsmath}        
\usepackage{nicefrac}       
\usepackage{microtype}      
\usepackage{graphicx}
\usepackage{natbib}
\usepackage{doi}
\usepackage{array}          
\usepackage{enumitem}       

\title{Who Sits Where? --- Automated Detection of Director Interlocks in
Indian Companies}

\author{ \hspace{1mm}Prateek Sancheti\\
    Data Sciences and Analytics Center\\
	International Institute of Information Technology\\
	Hyderabad, India \\
	\texttt{prateek.sancheti@research.iiit.ac.in} \\
	\And
    \hspace{1mm}Kamalakar Karlapalem\\
    Data Sciences and Analytics Center\\
	International Institute of Information Technology\\
	Hyderabad, India \\
	\texttt{kamal@iiit.ac.in} \\
    \And
    \hspace{1mm}Kavita Vemuri\\
    Biomedical Research Center\\
	International Institute of Information Technology\\
	Hyderabad, India \\
	\texttt{kvemuri@iiit.ac.in} \\
}

\renewcommand{\shorttitle}{Who Sits Where?}

\hypersetup{
  pdftitle={Who Sits Where? - Automated Detection of Director Interlocks in Indian Companies},
  pdfsubject={Corporate Governance, Network Analysis},
  pdfauthor={},
  pdfkeywords={Interlocking Directorships, Indian Corporations, LLMs, BFS graph traversal, Cliques},
  hidelinks,
}

\begin{document}
\maketitle

\begin{abstract}
Interlocking directorships---where individuals simultaneously serve on
the boards of multiple corporations---can facilitate the exchange of
expertise and strategic alignment but also present risks including
conflicts of interest, economic `oligarchy', and regulatory
non-compliance. In contexts such as large, family-controlled corporate
conglomerates in India, the manual detection of interlocks is hindered
by the high volume of corporate entities and the complex involvement of
extended familial networks. This study introduces a scalable,
graph-theoretic framework for the systematic identification and analysis
of interlocking directorships. Using Breadth-First Search (BFS)
traversal, we examined a curated dataset comprising over 50,000
directors, 85,000 companies, and 300,000 director--company affiliations,
yielding a comprehensive representation of corporate network structures.
Large Language Models (LLMs) were integrated into the analytical
pipeline to characterize both personal and professional linkages among
directors. Empirical results indicate that 17\% of directors hold
positions in exactly two companies, while 58.6\% maintain directorships
in two or more companies. The combined BFS--LLM methodology enables the
detection of recurrent director--company clusters, indicative of strong
network cohesion, and provides qualitative insights into potential
underlying drivers of these interlocks. The proposed approach enhances
the capacity for automated, data-driven detection of complex
intercorporate relationships, offering actionable implications for
corporate governance, regulatory monitoring, and systemic risk
assessment.
\end{abstract}

\keywords{Interlocking Directorships \and Indian Corporations \and LLMs
  \and BFS graph traversal \and Cliques}

\section{Introduction}

A corporation is a legal entity separate from its owners, capable of conducting business, owning assets, and being liable for its debts. These corporations are run by directors who are elected by shareholders to hold crucial responsibilities concerning the governance of the company, ensuring its operations align with shareholder interests. A corporation can have multiple individuals serving as directors, and
similarly, an individual can hold directorial positions at multiple corporations, creating a many-to-many network structure. In India as per the Companies Act 2013, a minimum of 2 and max of 15 directors are
allowed. \citet{jackling2009} testing for corporate governance by the
agency and resource dependency theories, found that larger board size
resulted in positive performance. This leads to the question of the
composition of the board of directors. In the complex and modern
structure of corporate governance, the phenomenon of interlocking
directorships (IDs) is a common yet nuanced aspect. An interlocking
directorship occurs when an individual holds directorial positions across
multiple corporations, creating a network of shared influence
\citep{mizruchi1996}. Such interlocks not only connect individuals with
diverse expertise and vision across corporate boards, but also foster an
intricate web of professional relationships that can have profound
implications for business practices, policy formation, and strategic
alliances. According to \citet{dooley1969}, the five key factors behind
the formation of board interlocks are: (1) corporation size, (2) degree
of managerial control, (3) financial ties, (4) interaction with rivals,
and (5) existence of local economic interests. Some studies also suggest
that resource-dependence theories provide another explanation for
interlocking \citep{pfeffer1972, pfeffer1978}. Early research on
interlocking directorates and corporate networks dates back to the
formative period of corporate capitalism in the early twentieth century
\citep{sapinski2018, dooley1969}. These studies focused on the reasons
for the creation and maintenance of interlocks
\citep{pfeffer1978, palmer1986, stearns1986}, whereas more recent
research has shifted attention to the consequences of such interlocks on
business behavior and corporate ideology
\citep{mizruchi1996, caiazza2019, hernandez2019}. While more recent
research has focused on impact on specific sectors like banking
\citep{barone2025}, policies like green innovation \citep{zhang2025},
gender inclusivity \citep{paoloni2025} and family involvement and IPO
pricing \citep{singh2025}.

The dual-edged nature of director interlocking raises significant
concerns. On one hand, engaging visionary leaders and implementing strong
corporate governance can promote the dissemination of innovative policies
and allow for the flow of valuable information and resources. Interlocks
can also help corporations gain a competitive edge by monitoring business
ties and gathering confidential information, which leads to improved
management practices
\citep{loderer2002, ozmel2013, lamb2016, mazzola2016} and innovation
\citep{helmers2017}. On the other hand, the independence of board
decisions might be compromised by shared directors \citep{adams2017},
harbor potential risks related to corporate malpractices or quid pro quo
situations, including, but not limited to, corruption, collusion, the
formation of shell companies, facilitation of hostile takeovers, and
unfair business practices \citep{hillman1999, holburn2008} and can form
an economic oligarchy.

Transparency and awareness among investors, stakeholders, and regulatory
bodies regarding such interconnected networks are crucial for maintaining
the integrity of corporate governance. Motivated by the critical need for
clearer insights into the structure of director interlocking, particularly
within the context of Indian corporations, we present a pipeline to
navigate through the complexities of these networks. We identify weakly
and strongly connected components using Graph Cliques and Frequent
Itemsets of directors and companies within these corporate networks,
highlighting how a small set of directors can control many companies and
how a small set of companies can be controlled by a large number of
directors. Additionally, we identify relationships among connected
directors by extracting director profiles from the web. These
relationships are categorized as either personal, including family ties
such as Daughter--Father or Wife--Husband, or professional, encompassing
shared work experiences, professional memberships, or educational
backgrounds. The objective is to provide nuanced information and model
for an informed analysis of corporate structures.

The contribution of the paper is the method to mine, retrieve and analyse
large web data of corporates and extract the interlocks. Towards this we
leverage a BFS graph traversal-based approach. We extracted a data corpus
of over 50,000 directors, 85,000 companies, and 300,000 company-director
connections. We analyze the corporate networks and present results
revealing insights into the complex corporate structures of Indian
corporations.

\subsection{Cliques, Maximal Cliques, and Maximum Cliques}

In the context of graph theory, a \textit{Clique} represents a specific
subset of vertices within an undirected graph, characterized by complete
adjacency among its members. A \textit{Maximal Clique} is a clique that
cannot be expanded by including any additional adjacent vertices. This
definition implies that a maximal clique is not wholly contained within
the vertex set of a larger clique. Essentially, it represents a subset of
vertices that forms a complete graph and is as large as possible without
being subsumed by a greater clique. This characteristic makes maximal
cliques particularly significant in network analysis, as they help
identify the largest connected groups.

Mathematically and computationally, identifying all cliques of all
maximal is a NP-Complete Problem \citep{hua2020}. To extract all the
cliques in a graph $G$ with vertices $V$, we need to identify all subsets
and check if all the vertices in the subgraph are connected to each
other. For a subgraph with $n$ nodes, to be a clique needs $\binom{n}{2}$
edges and therefore the time for extracting all maximal cliques in the
graph $G$ can be exponential time of the vertex number $V$
\citep{moon1965}. Various approaches such as \textit{Pivot
Bron-Kerbosch Algorithm} and \textit{Tomita Algorithm} have been studied
to identify number of Cliques in a given graph.

\subsection{Frequent Itemsets and Maximal Frequent Itemsets}

In data mining, the concept of \textit{Frequent Itemsets}, often related
with association rule mining, is used to extract relationships among items
within a dataset. A \textit{Frequent Itemset} consists of a combination
of items that appear together with considerable regularity in the dataset,
quantified by a support count. This count reflects the number of
transactions or records in which the itemset occurs.

A \textit{Maximal Frequent Itemset} is defined as a frequent itemset that
cannot be enlarged by adding another element without losing its frequent
status in the dataset. In other words, none of its immediate supersets
appear frequently enough to meet the minimum support criterion, making
these itemsets a concise and compact representation of all frequent
itemsets within the dataset. Various algorithms, such as \textit{Apriori
or FP-Growth}, are utilized to extract frequent itemsets. We use the
FP-Growth Algorithm's implementation.

\subsection{Frequent Pattern Growth (FP-Growth) Algorithm}

The \textit{Frequent Pattern Growth (FP-Growth) Algorithm}, developed by
\citet{han2000}, is another widely used algorithm in the field of data
mining, renowned for its superior efficiency in extracting frequent
itemsets from large-scale datasets. Central to this algorithm is the
construction of a \textit{Frequent Pattern Tree (FP-Tree)}, which uses a
tree-like data structure to represent the dataset in a compact yet
effective manner, capturing the essential transaction information. The
process of FP-Tree construction begins by reading each transaction and
mapping it onto a specific path in the data structure. Once the FP-Tree
is fully constructed, the algorithm generates frequent itemsets by
recursively mining the tree. This mining process starts at the bottom of
the tree and proceeds upwards in a bottom-up manner. It systematically
explores all possible combinations of itemsets, identifying those that
meet or exceed the minimum support threshold.

\section{Methodology}

The primary data source for our pipeline is a publicly available site
that features two main categories of web pages: Company Pages and
Director Pages. A Company Page lists basic information about the company,
including the company name, Company Identification Number (CIN), company
address, email IDs of point-of-contact individuals, and hyperlinks to
current and past directors. A Director Page contains details about the
directors, such as their name, Director Identification Number (DIN), and
links to all companies where they hold or have held directorial
positions. This structure allows us to conceptualize a potentially
endless network of corporate connections as a graph.

All individuals on the board of directors of a company, including
individual partners, nominee-body partners, directors, managing
directors, partners, etc., are considered as Director Nodes. A corporate
network spanning companies and their directors is a two-mode network,
where nodes can be classified into two categories, with edges connecting
categories. We employ a \textit{Breadth-First Search (BFS)} graph
traversal technique to explore the corporate network. The data extraction
process begins with the user providing a base node, which serves as the
starting vertex for the graph traversal. The BFS technique, aided by a
\textit{queue} and a list of \textit{visited nodes}, ensures that no
nodes are revisited. This extracted data forms the foundation for a
detailed examination and analysis of the interlocking directorates
network in the subsequent steps of our methodology. The extracted
information about the corporate network is organized into three distinct
files:

\begin{enumerate}
  \item Company Information: List of all the companies in the entire
    network with all the relevant information --- \textit{CIN, Company
    Name, URL}.
  \item Director Information: List of all the directors in the entire
    network with all the relevant information --- \textit{DIN, Director
    Name, URL}.
  \item Company -- Director Information: A list of all companies and
    their directors. Each row has a company name and a director name
    implying directorship of an individual at that company.
\end{enumerate}

Post extraction of the data and completion of the analysis, company name,
individual name or any other form of identification is anonymized.

\subsection{Data Visualization}

There are five distinct types of visualizations that can be effectively
used for a corporate network, each designed to highlight different aspects
and patterns within the data.

\textbf{Company-Director Graph:} This is a foundational visualization
derived from the dataset. It represents a network structure in which both
companies and directors are nodes. Each connection, or edge, between a
director and a company signifies a directorship.

\begin{figure}[ht]
  \centering
  \includegraphics[width=0.4\textwidth]{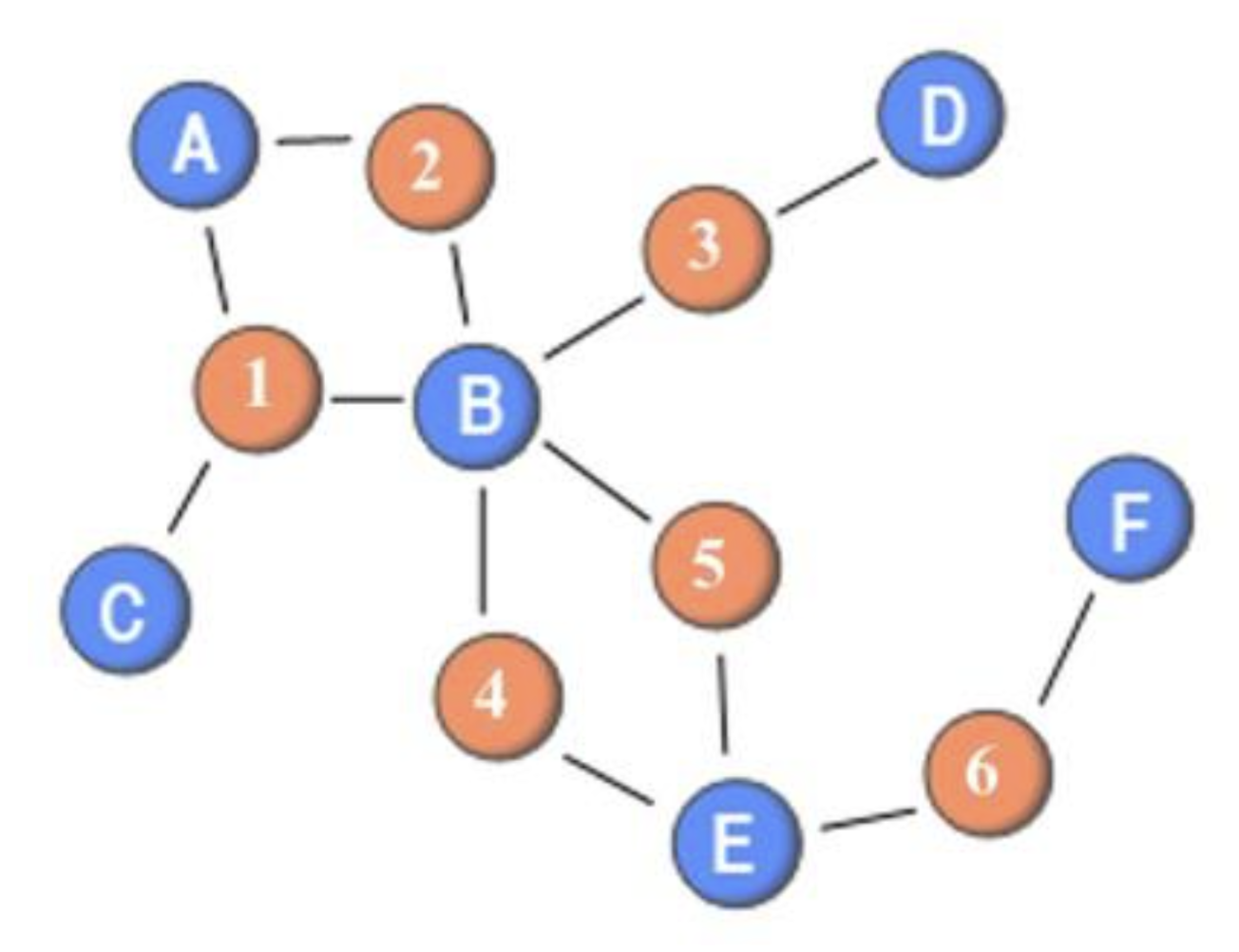}
  \caption{\centering Company -- Director Graph: Blue Nodes (A--F) depict Companies, Orange Nodes (1--6) depict Directors.}
  \label{fig:cd-graph}
\end{figure}

In Figure~\ref{fig:cd-graph}, the blue nodes represent companies, and
the orange nodes represent directors. Company nodes, such as `B', connect
to the majority of the nodes in the network. We refer to such nodes,
which have a large number of directors, as \textit{Star Company Nodes}.
If such a node ceases to exist for any reason, such as insolvency, it
would cause the entire network to disconnect. A director node that
connects to many other directors through shared directorships or serves
on the boards of many companies is called a \textit{Star Director Node}.
In the event of a director's death or retirement, such a node would
disconnect the entire network.

\subsubsection{Company-Company Graphs (Direct \& Indirect)}

The Company-Company Graphs offer a distinct perspective, focusing solely
on the direct relationships between companies within the dataset. This is
a one-mode graph in which companies are the only nodes. In the Direct
Company-Company graph, two companies are connected if they share a
director, with the edge signifying the shared directorial tie. In an
Indirect Company-Company Graph, two companies are connected if a path
exists between them in the two-mode representation of the network.

\begin{figure}[ht]
  \centering
  \includegraphics[width=0.46\textwidth]{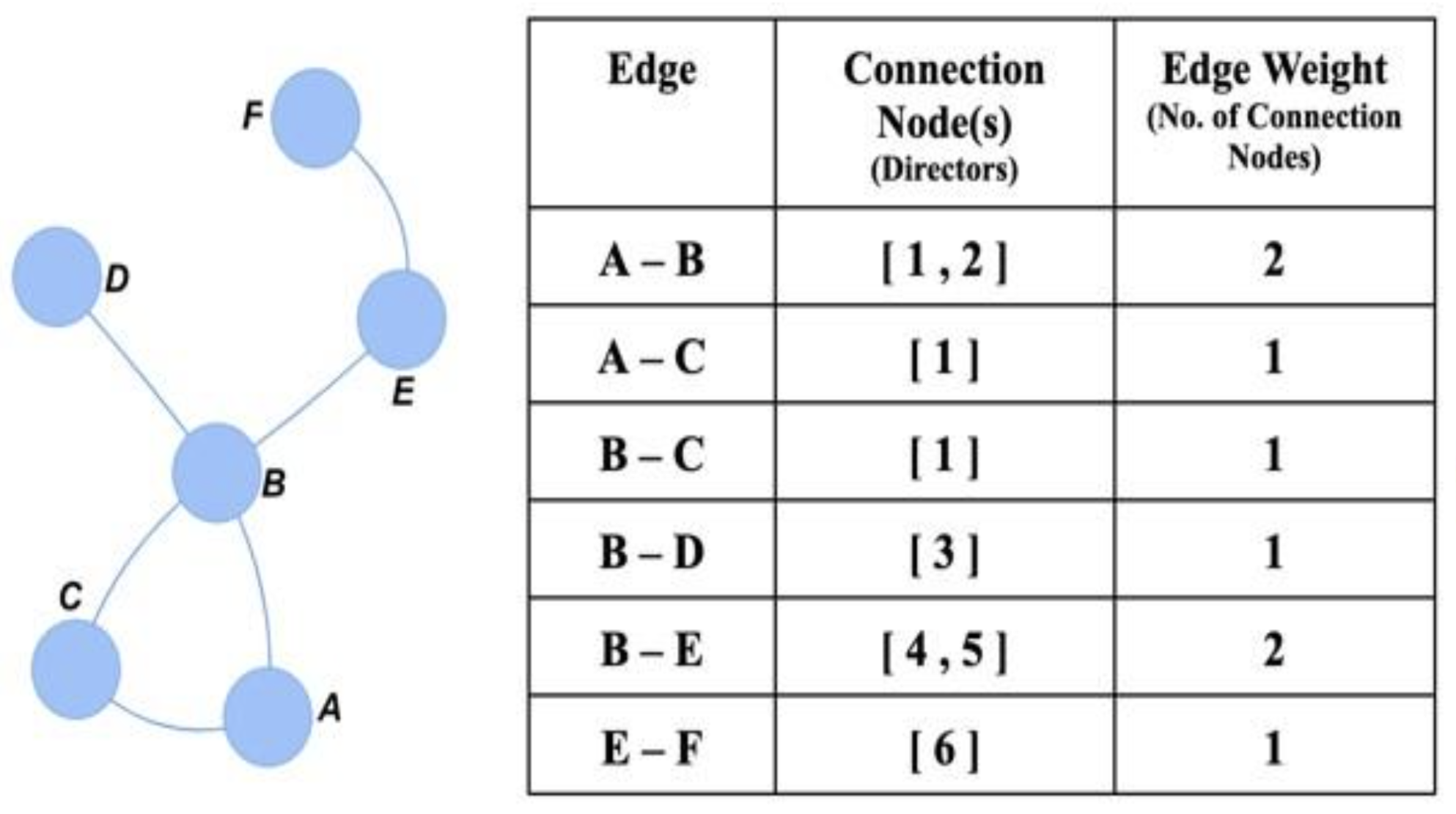}
  \includegraphics[width=0.48\textwidth]{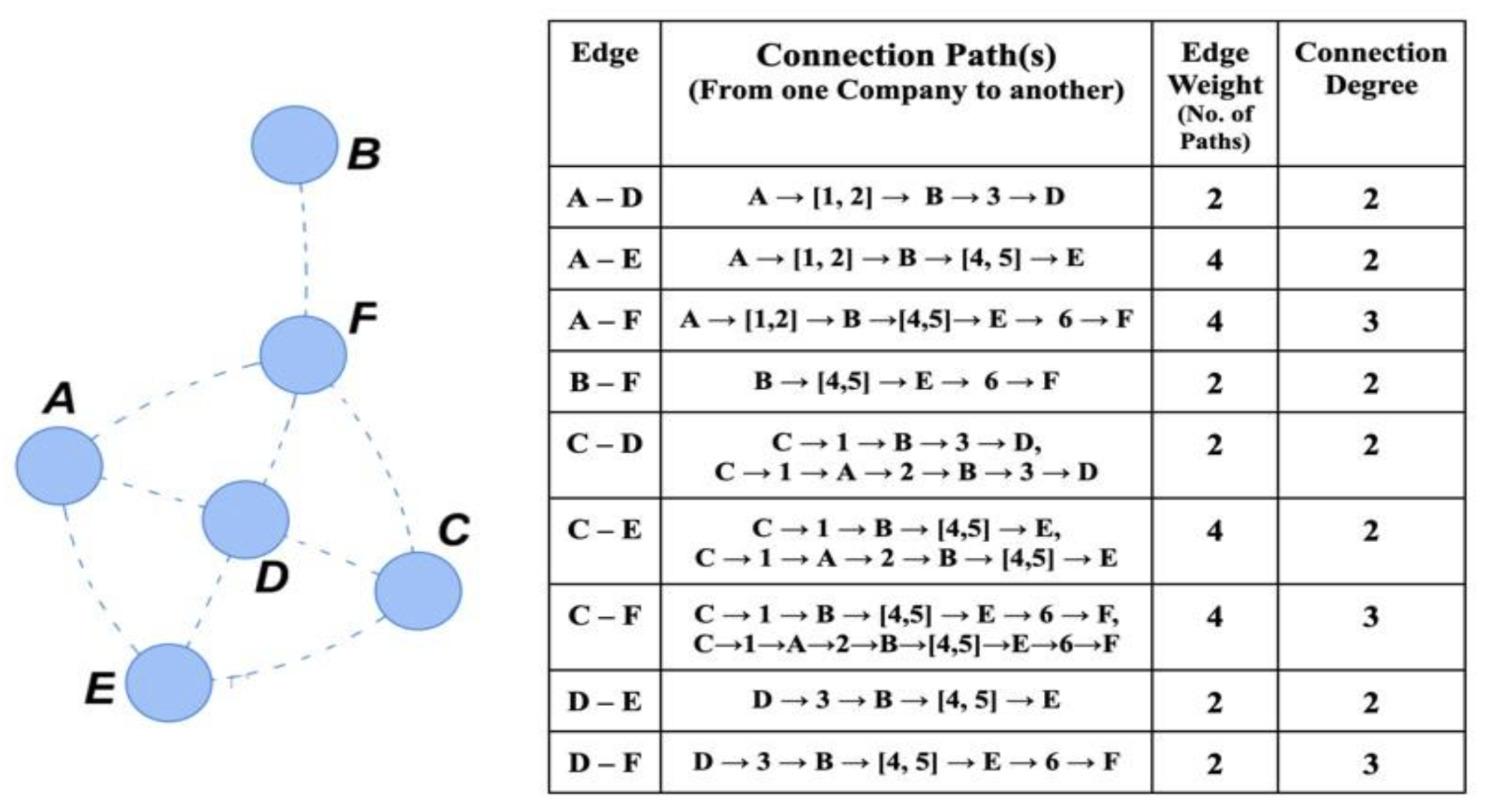}
  \caption{\centering Company -- Company Graphs (Direct and Indirect). (a) Direct
    Company -- Company Graph\quad (b) Indirect Company -- Company Graph.}
  \label{fig:cc-graph}
\end{figure}

The visualization help gain insights into the web of interdependencies
among companies, highlighting potential areas of collaboration,
competition, or shared strategic interests.
Figure~\ref{fig:cc-graph}(a) and Figure~\ref{fig:cc-graph}(b) illustrate
the Direct and Indirect Company-Company Graphs, respectively, derived
from the corporate network shown in Figure~\ref{fig:cd-graph}. The
accompanying table in Figure~\ref{fig:cc-graph}(a) provides a detailed
account of each edge, elucidating the rationale behind the
connections/edges between pairs of companies, as well as the strength of
each connection through the Edge Weight. A higher edge weight indicates a
greater number of shared directors and thus a stronger connection between
two companies. Companies with a shared director are referred to as
\textit{1st Degree} connections. The accompanying table in
Figure~\ref{fig:cc-graph}(b) offers a comprehensive breakdown of each
indirect connection observed in the network by describing all possible
connection paths for each pair. For example, Company Nodes `A' and `D'
do not share a director, but there are two different paths between them:
(i) A $\rightarrow$ 1 $\rightarrow$ B $\rightarrow$ 3 $\rightarrow$ D
and (ii) A $\rightarrow$ 2 $\rightarrow$ B $\rightarrow$ 3 $\rightarrow$
D. The connection degree of an edge here is the number of distinct
directors present along the shortest path between the two companies,
providing insight into how far apart the two company nodes are.

It is interesting to notice how Company Node `B' is common to all the
paths for indirectly connected companies in Figure~\ref{fig:cc-graph}(b).
This highlights how Company `B' plays a pivotal role in connecting
several companies both directly and indirectly within the network.
Another important observation is the Edge Weight, or the number of paths
between two companies. Given two pairs of companies that are indirectly
connected and have the same connection degree, the pair with more paths
between them will be considered more strongly connected than the pair
with fewer paths. For instance, company pairs `A' -- `D' and `A' -- `E'
are both \textit{2nd Degree} connections, but they have a total of 2 and
4 paths, respectively. Hence, company `A' can be said to be more
connected to company `E' than to company `D'.

\subsubsection{Director-Director Graph (Direct \& Indirect)}

\begin{figure}[ht]
  \centering
  \includegraphics[width=0.41\textwidth]{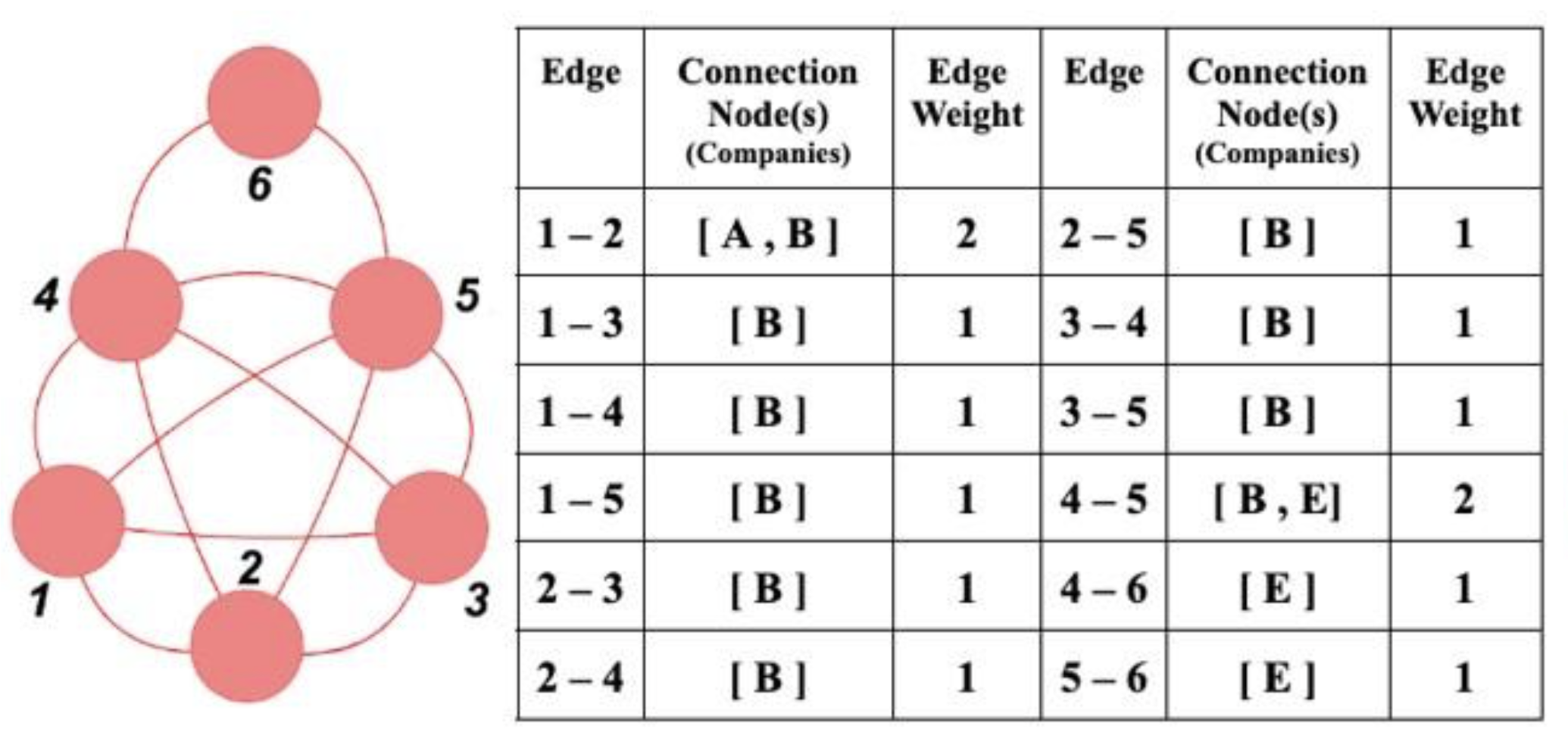}
  \includegraphics[width=0.49\textwidth]{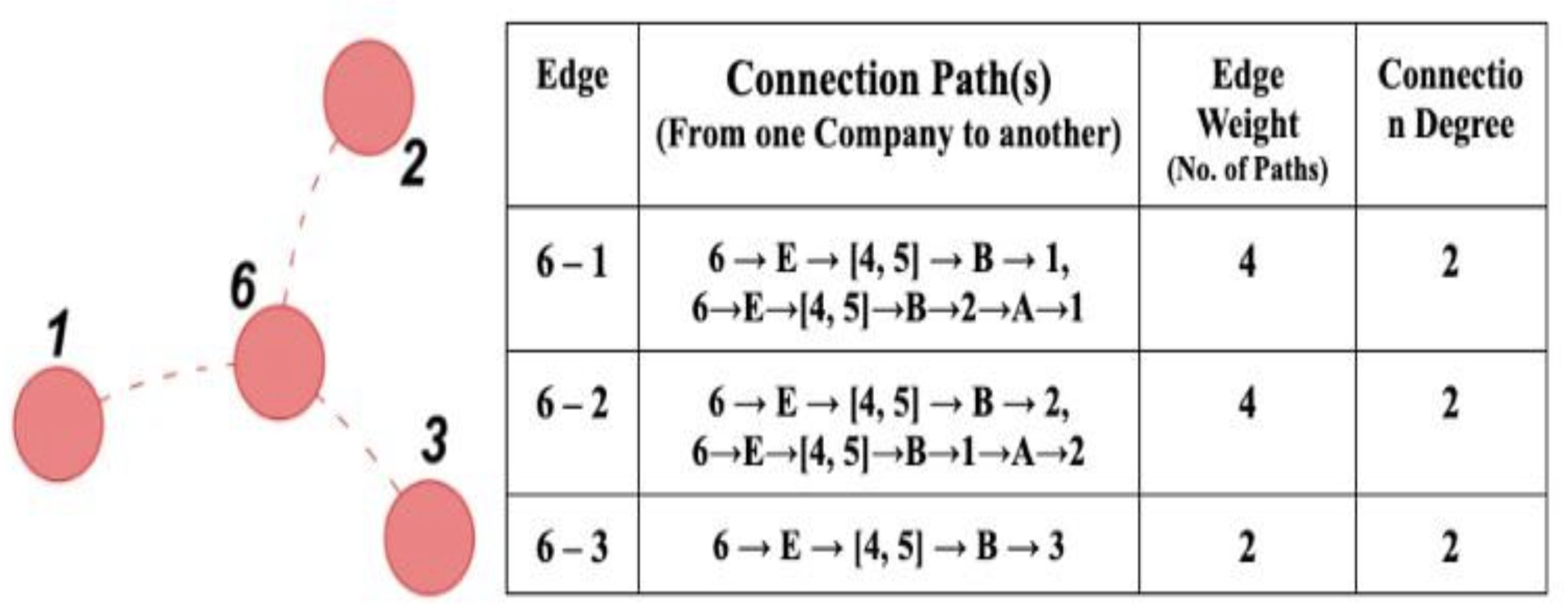}
  \caption{\centering Director -- Director Graphs (Direct and Indirect). (a) Direct
    Director -- Director Graph\quad (b) Indirect Director -- Director
    Graph.}
  \label{fig:dd-graph}
\end{figure}

A Director-Director Graph mirrors the structure of the Company-Company
Graphs but focuses on individuals. In a Direct Director-Director Graph,
two directors are connected if they share directorships at any company.
In contrast, an Indirect Director-Director Graph connects two directors
if a path exists between them. Figure~\ref{fig:dd-graph}(a) and
Figure~\ref{fig:dd-graph}(b) serve as illustrative depictions of the
Direct and Indirect Director-Director Graphs, respectively,
corresponding to the network in Figure~\ref{fig:cd-graph}.

The accompanying table in Figure~\ref{fig:dd-graph}(a) elucidates the
directorial ties between pairs of individuals, as well as the strength
of each connection through Edge Weight. A higher edge weight indicates a
greater number of shared companies, and thus a stronger connection
between two directors. Directors who share a company can be referred to
as \textit{1st Degree} connections. The accompanying table in
Figure~\ref{fig:dd-graph}(b) describes each indirect connection observed
between directors in the network, along with all the possible connection
paths for each pair. For example, Director Nodes `1' and `6' do not
share a company, but there are 4 different paths between them: (i)
$6 \rightarrow E \rightarrow 4 \rightarrow B \rightarrow 1$, (ii)
$6 \rightarrow E \rightarrow 5 \rightarrow B \rightarrow 1$, (iii)
$6 \rightarrow E \rightarrow 4 \rightarrow B \rightarrow 2 \rightarrow
A \rightarrow 1$, and (iv) $6 \rightarrow E \rightarrow 5 \rightarrow B
\rightarrow 2 \rightarrow A \rightarrow 1$. The connection degree of an
edge here is the number of distinct companies present along the shortest
path between the two directors, providing insight into how far apart
the two director nodes are.

Similar to the Company-Company Indirect Graph, given two pairs of
directors that are indirectly connected and have the same connection
degree, the pair with more paths between them will be considered more
strongly connected than the pair with fewer paths. For instance, the
director pairs 1 -- 6, 2 -- 6, and 3 -- E are all \textit{2nd Degree}
connections, with 4, 4, and 2 paths, respectively. Hence, Director `6'
can be said to be more connected with Directors `1' and `2' than with
Director `3'. What makes the Director-Director Graph particularly
significant is its focus on the human factor within corporate networks.

\subsection{Maximal Cliques in Corporate Networks}

Cliques and Maximal Cliques assist in analyzing interconnectivity and
network robustness. Our focus is on identifying maximal cliques in a
given network, as this helps extract large groups of companies and
directors connected to each other. To address the computational
complexity of extracting all maximal cliques present in a network while
still gaining meaningful insights, we limit the network analysis to
third-degree connections of a specific node (either a director or a
company). This effectively creates a smaller, more manageable subnetwork,
allowing for the identification of potentially influential communities
within the immediate vicinity of the chosen node.

\begin{figure}[ht]
  \centering
  \includegraphics[width=0.22\textwidth]{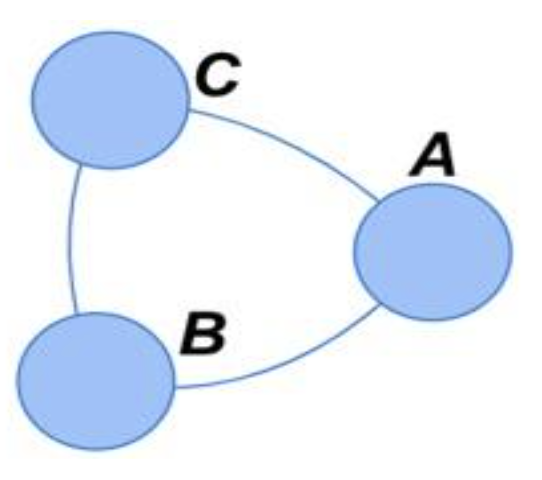}
  \includegraphics[width=0.45\textwidth]{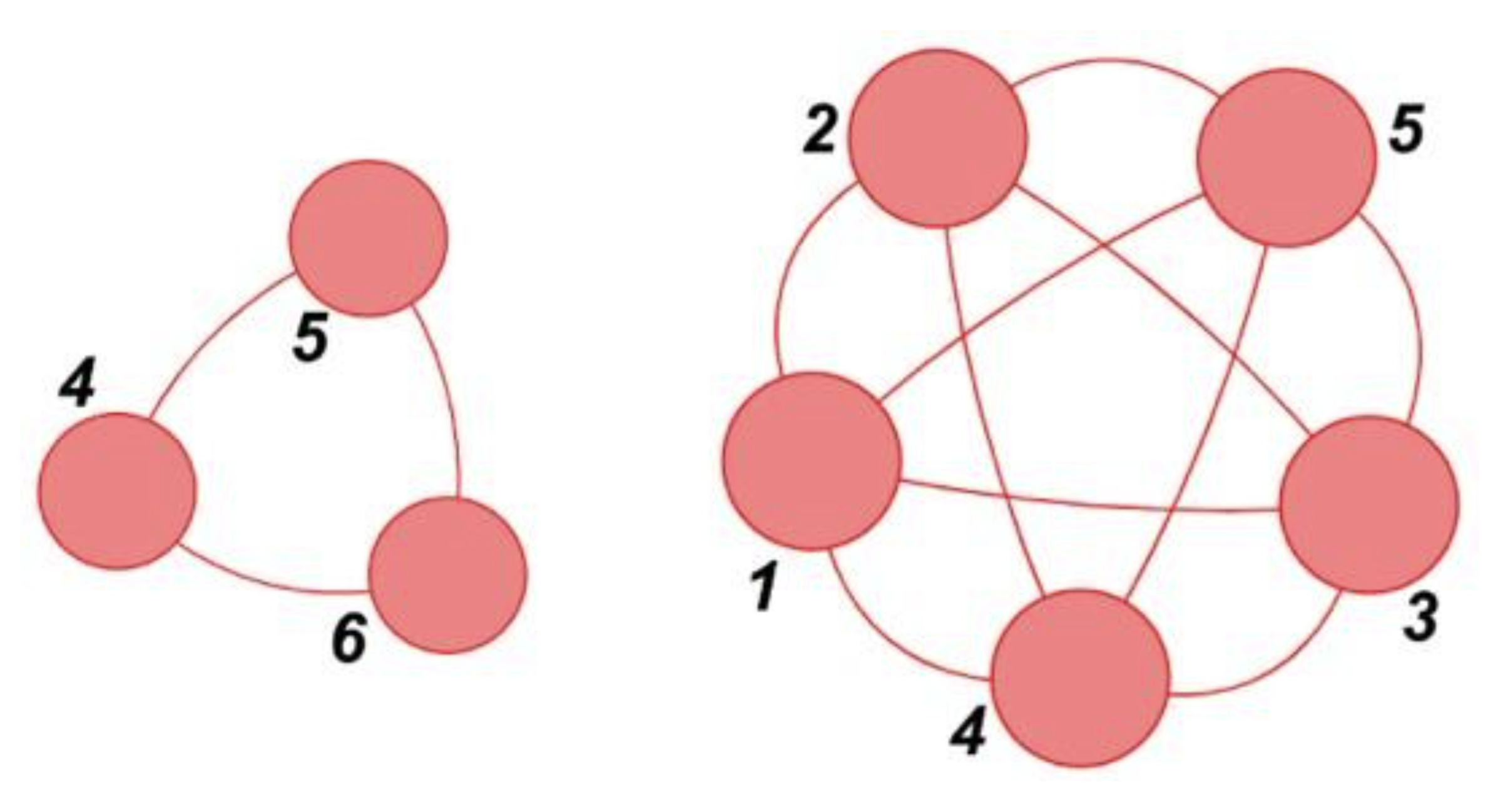}
  \caption{\centering (a) Maximal Company Cliques for Figure~\ref{fig:cc-graph} (a) - 3-Vertex Clique (ABC)\newline(b) Maximal Director Cliques for Figure~\ref{fig:dd-graph}(a) - 5-Vertex Clique (12345)}
  \label{fig:cliques}
\end{figure}

We use the \textit{find\_cliques} function from the \textit{NetworkX
library} in Python. The implementation of this function is based on the
discussion by \citet{cazals2008} on the Bron and Kerbosch algorithm
\citep{bron1973}, along with its extension by \citet{tomita2004}.
Figure~\ref{fig:cliques} shows the maximal cliques for the sample network
of directors and companies discussed earlier in
Figure~\ref{fig:cd-graph}. Various cliques (size $> 2$) can be
identified for the corresponding Company-Company Direct Graph
(Figure~\ref{fig:cc-graph}(a)) and Director-Director Direct Graph
(Figure~\ref{fig:dd-graph}(a)).

\textbf{Maximal Company Cliques}\\
The network in Figure~\ref{fig:cc-graph}(a) contains only one maximal
company clique, as depicted in Figure~\ref{fig:cliques}(a). Companies A,
B, and C form a maximal clique of size 3. This can be interpreted as
meaning that all the companies in this maximal clique are connected and
thus have at least one director in common for each pair of companies.
Companies A and B share two directors---1 and 2---whereas Companies A
and C, as well as Companies B and C, share only one director---1.

\textbf{Maximal Director Cliques}

The network in Figure~\ref{fig:dd-graph}(a) contains two maximal
director cliques, as depicted in Figure~\ref{fig:cliques}(b). Directors
1, 2, 3, 4, and 5 form a maximal clique of size 5, and Directors 4, 5,
and 6 form a maximal clique of size 3. This can be interpreted as meaning
that all the directors in each maximal clique are connected and thus have
at least one company in common for each pair of directors. Directors 1
and 2, and Directors 4 and 5, share two directors---A, B and B, E,
respectively. All the other pairs of directors share one director---B.

A larger clique, or a clique with more nodes (directors or companies),
has more power to create influence in a corporate network compared to a
smaller clique. For instance, in Figure~\ref{fig:cliques}(b), the
5-vertex director clique (\textit{12345}) is larger in size than the
3-vertex director clique (\textit{456}). With more directors connected to
each other, it has more power to create interlocking influence.

Although every pair of directors or companies in a clique is connected,
the level or strength of each connection differs. We cannot assume that
all pairs are equally connected. A pair of directors that share more
companies is more strongly connected than a pair that shares fewer
companies. Similarly, this applies to company cliques. Hence, maximal
cliques in a corporate network can be viewed as weakly connected
components. To identify strongly connected components, we turn to mining
frequent itemsets of directors and companies in the network.

\subsection{Maximal Frequent Itemsets in Corporate Networks}

To effectively identify maximal frequent itemsets within a corporate
network---such as groups of directors or companies that frequently appear
together---the extracted data is reorganized into specific formats
required for this analysis. For each company in the Company Information
file, a list of directors holding directorial positions is prepared.
Similarly, for each director in the Director Information file, a list of
companies where the individual is a director is compiled.

We use the \textit{fpmax} function developed by the \textit{mlxtend
library} to extract maximal frequent itemsets from a database. The
implementation of this function is based on the algorithm developed by
\citet{grahne2003}. The function takes in \textit{min\_support} as an
input, which acts as the threshold for an item to be considered frequent.
For example, if the dataset contains 10,000 rows and the
\textit{min\_support} is 0.0001, an itemset must occur at least
$10{,}000 \times 0.0001 = 10$ times to be considered frequent.

\subsubsection{Maximal Frequent Director Itemsets (MFDIs)}

We identify sets of directors that frequently appear together in the
corporate network. These are the strongly connected directors in the
network. We also identify the companies corresponding to each frequent
director itemset. To extract this, we use the list of directors
corresponding to each company prepared earlier. Table~\ref{tab:mfdis}
shows an example of MFDIs, with the support denoting the number of times
the corresponding itemset has appeared in the dataset. The higher the
support, the stronger the itemset. The interesting companies
corresponding to an itemset represent the companies where all the
individuals in the itemset serve as directors.

\begin{table}[h]
  \caption{Example Results Maximal Frequent Director Itemset (MFDIs) and
    Corresponding Companies}
  \label{tab:mfdis}
  \centering
  \begin{tabular}{lll}
    \toprule
    \textbf{Support / Frequency} & \textbf{Frequency Director Itemset}
      & \textbf{Intersecting Companies} \\
    \midrule
    $s_1$ & $[D_4\ \ldots]$        & $[C_x\ \ldots]$ \\
    $s_2$ & $[D_4, D_7, D_5\ \ldots]$ & $[C_a\ \ldots]$ \\
    $s_3$ & $[D_2, D_3\ \ldots]$   & $[C_x, C_a, C_z\ \ldots]$ \\
    \ldots & \ldots                 & \ldots \\
    \bottomrule
  \end{tabular}
\end{table}

\subsubsection{Maximal Frequent Company Itemsets (MFCIs)}

We identify sets of companies that frequently appear together, sharing
directors, in a corporate network by utilizing the list of companies
corresponding to each director. These are the strongly connected
companies in our network. We also identify the directors corresponding
to these frequent company itemsets. Table~\ref{tab:mfcis} shows examples
of MFCIs. \textit{Support} denotes the number of times the corresponding
itemset has appeared in the dataset of directors. The higher the support,
the stronger the itemset. The interesting directors corresponding to an
itemset are the individuals who serve as directors at all the companies
in the itemset.

\begin{table}[h]
  \caption{Example Results Maximal Frequent Company Itemset (MFCIs) and
    Corresponding Directors}
  \label{tab:mfcis}
  \centering
  \begin{tabular}{lll}
    \toprule
    \textbf{Support / Frequency} & \textbf{Frequency Company Itemset}
      & \textbf{Intersecting Directors} \\
    \midrule
    $s_1$ & $[C_y\ \ldots]$           & $[D_1\ \ldots]$ \\
    $s_2$ & $[C_p, C_q, C_r\ \ldots]$ & $[D_3\ \ldots]$ \\
    $s_3$ & $[C_z, C_x\ \ldots]$      & $[D_2, D_1, D_5\ \ldots]$ \\
    \ldots & \ldots                   & \ldots \\
    \bottomrule
  \end{tabular}
\end{table}

\subsection{Data Description}

A comprehensive network of corporate directors and the companies they
are affiliated with was built for this study using the data extraction
technique explained earlier. Since we need to provide a base node to
begin graph traversal, as described previously, we tested the approach on
a large conglomerate: \textit{ABCD} (anonymised but represents one of the
largest company in India). The dataset comprises a total of 54,216
directors, 87,187 companies, and 299,970 company-director edges.

\begin{figure}[ht]
  \centering
  \includegraphics[width=0.45\textwidth]{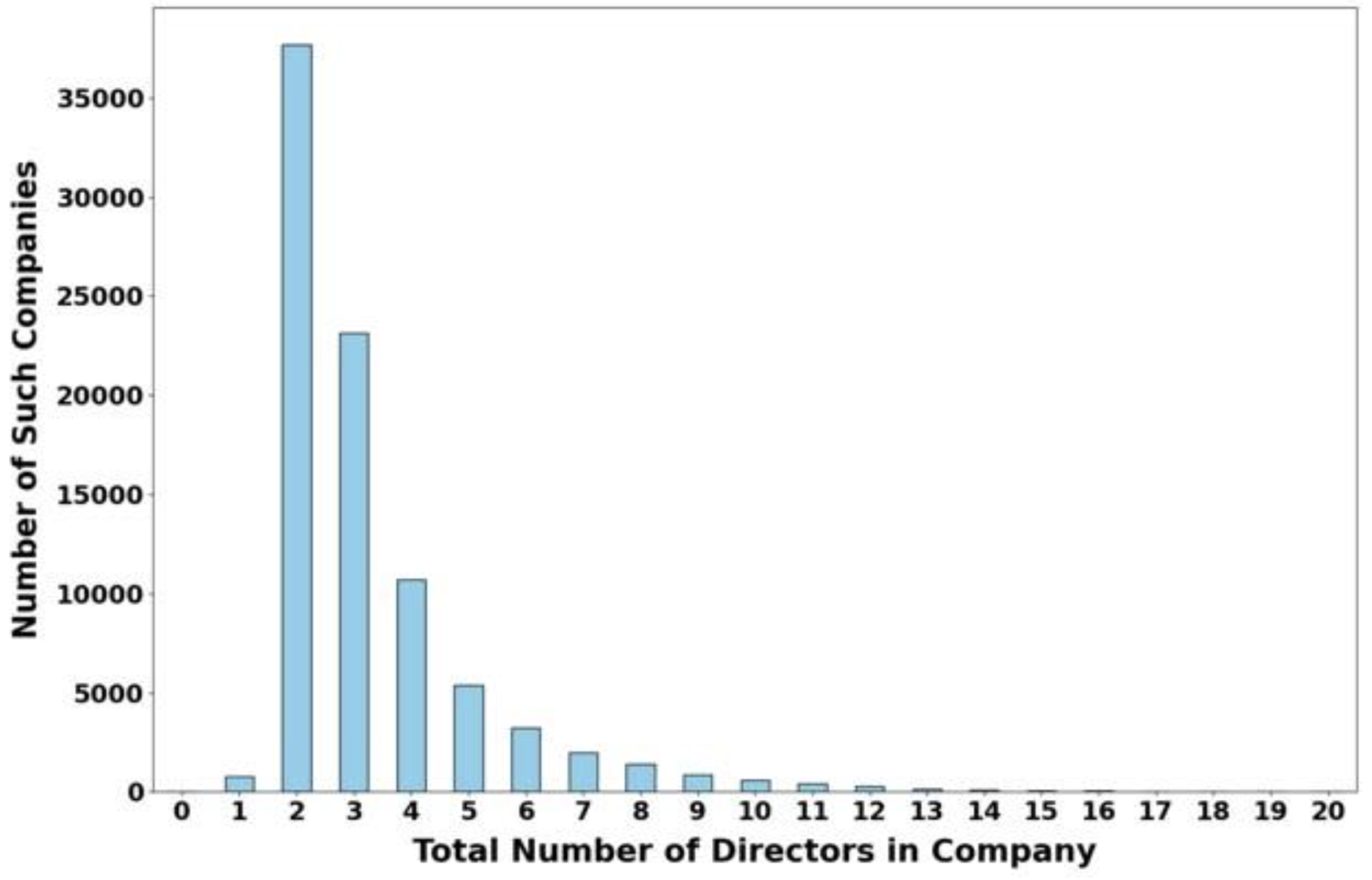}
  \includegraphics[width=0.45\textwidth]{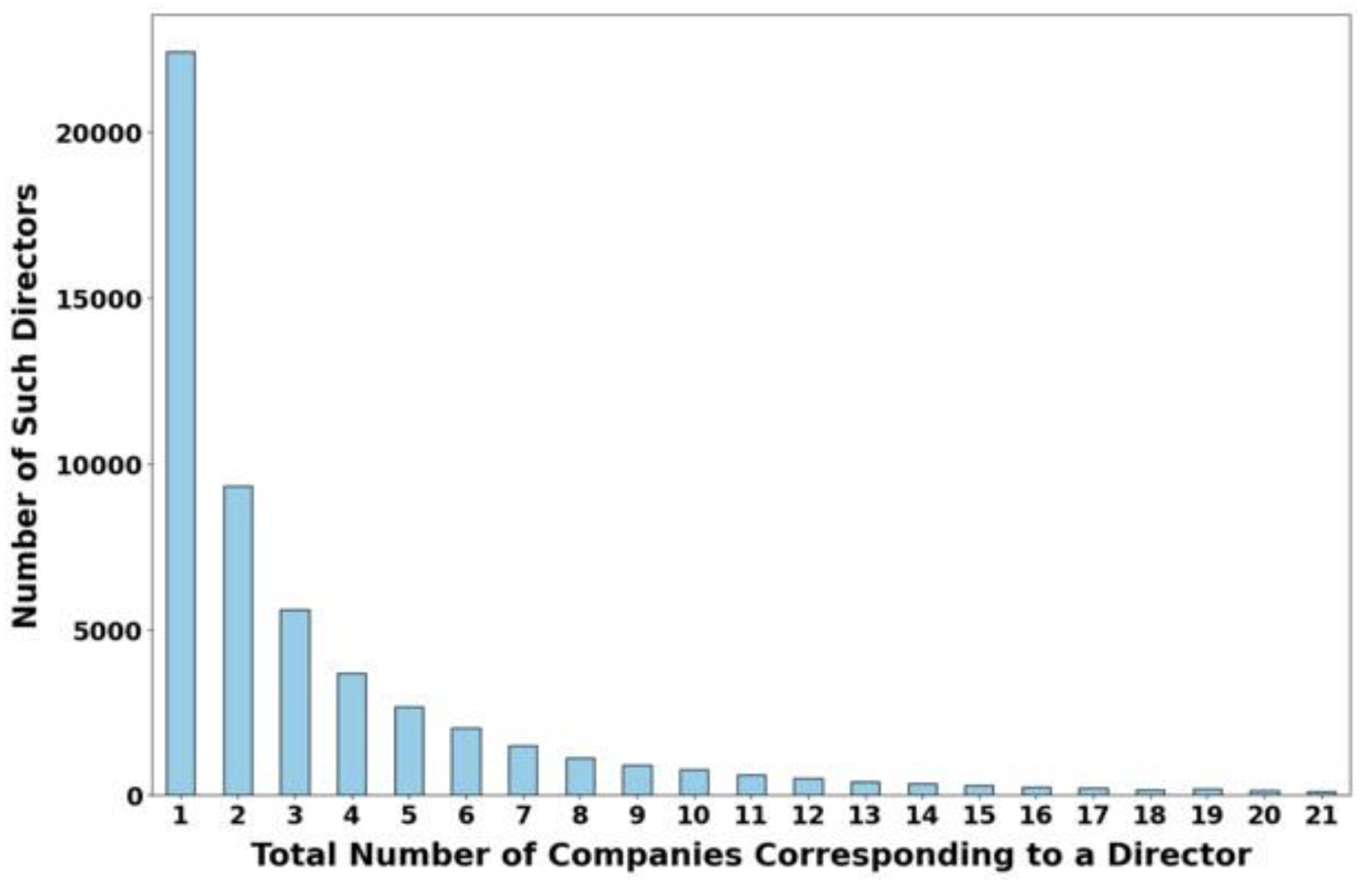}
  \caption{\centering (a) Frequency of Number of Directors Corresponding to a Company\quad (b) Frequency of Number of Companies Corresponding to a
    Director.}
  \label{fig:distributions}
\end{figure}

Figure~\ref{fig:distributions}(a) represents the frequency distribution
of the number of directors for a company. We observe that most companies
have either 2 directors (43\% of the total companies) or 3 directors
(26\% of the total companies). Figure~\ref{fig:distributions}(b), on the
other hand, shows the frequency distribution of the number of companies
corresponding to a given individual (director). We observe that most
individuals hold directorial positions at either 1 company (41\% of the
total directors) or 2 companies (17\% of the total directors).

Observations: From one base node, the resulting network of directors and
companies includes business heads of other large conglomerates
(\textit{anonymized}) represented as \textit{AA, BB, CC, DD, EE, FF, GG,
HH}. A few of these conglomerates such as the KK, AA, BB and MM, are 1st
degree connections to each other with some companies sharing directors,
such as:

\begin{itemize}
  \item \textit{X1}: KK -- AA
  \item \textit{X2}: KK -- BB
  \item \textit{X3}: KK -- NN
  \item \textit{X4}: KK -- PP
  \item \textit{X5}: QQ. -- PP etc.
\end{itemize}

\noindent Where `X1'--`X5' are director IDs, while KK, AA, BB, NN, PP,
QQ, PP represent conglomerates.

\section{Results}

\subsection{Maximal Director Cliques (MDCs)}

It is not feasible to extract all possible cliques from the entire
corporate network that we scraped. Therefore, to illustrate the utility
of identifying maximal director cliques within a corporate network, we
conducted an analysis on five example cases and a few directors (X6-10)
of their own large business. For each base director, we extracted the
first, second, and third-degree connections (directors).

\begin{itemize}
  \item Cliques observed among first and second-degree connections will
    be a subset of those observed among third-degree connections, so we
    focus on extracting cliques for third-degree connections.
  \item We extract all MDCs that include the base director.
  \item The statistics are presented after sorting these MDCs based on
    two criteria: (1) the number of directors in an MDC, and (2) the
    number of intersecting companies where these directors share board
    seats.
\end{itemize}

Table~\ref{tab:mdc} presents a consolidated representation of the
results for the example MDCs, where:

\begin{itemize}
  \item Col A: Number of directors in 3rd-degree connections, Number of
    companies in 3rd-degree connections
  \item Col B: Number of MDCs that include the corresponding director,
    Average number of directors in such MDCs
  \item Col C: Size of the MDC with the maximum number of directors,
    Number of companies shared
  \item Col D: Size of the MDC with the minimum number of directors,
    Number of companies shared
  \item Col E: Size of the MDC that shares the maximum number of
    companies, Number of companies shared
  \item Col F: Size of the MDC that shares the minimum number of
    companies, Number of companies shared
\end{itemize}

\begin{table}[h]
  \caption{Consolidated Results for MDCs Examples.}
  \label{tab:mdc}
  \centering
  \begin{tabular}{lllllll}
    \toprule
    \textbf{Director Name} & \textbf{A} & \textbf{B} & \textbf{C}
      & \textbf{D} & \textbf{E} & \textbf{F} \\
    \midrule
    \textit{X6}  & 2996, 1100  & 4, 4.75  & 6, 3  & 2, 3 & 7, 4  & 3, 1 \\
    \textit{X7}  & 6102, 2960  & 17, 8.05 & 12, 10 & 4, 5 & 9, 16 & 8, 2 \\
    \textit{X8}  & 9178, 4002  & 22, 5.45 & 9, 8  & 3, 3 & 5, 21 & 3, 3 \\
    \textit{X9}  & 8272, 4037  & 16, 6.31 & 12, 4 & 3, 3 & 8, 9  & 5, 1 \\
    \textit{X10} & 7151, 3452  & 6, 7.5   & 14, 7 & 2, 1 & 10, 13 & 2, 1 \\
    \bottomrule
  \end{tabular}
\end{table}

Observations (Table~\ref{tab:mdc}):

\begin{itemize}
  \item For two MDCs with the same number of directors, the higher the
    number of shared companies, the stronger the connection between
    directors.
  \item Similarly, for two MDCs sharing the same number of companies,
    the fewer the number of directors in the MDC, the stronger the
    connection between directors.
  \item A stronger MDC can exert more influence compared to a weaker MDC.
\end{itemize}

\subsection{Maximal Company Cliques (MCCs)}

We conducted an analysis to extract MCCs for five companies:

\begin{itemize}
  \item We extracted the first, second, and third-degree connections
    (companies) for each base company.
  \item We extracted all MCCs in third-degree connections where the base
    company is present.
  \item The statistics are presented after sorting these MCCs based on
    two criteria: (1) the number of directors in an MCC, and (2) the
    number of intersecting companies where these directors share board
    seats.
\end{itemize}

Table~\ref{tab:mcc} presents a consolidated representation of the
results for the example MCCs, where:

\begin{itemize}
  \item Col P: Number of companies in 3rd-degree connections, Number of
    directors in 3rd-degree connections
  \item Col Q: Number of MCCs that include the corresponding company,
    Average number of companies in such MCCs
  \item Col R: Size of the MCC with the maximum number of companies,
    Number of directors shared
  \item Col S: Size of the MCC with the minimum number of companies,
    Number of directors shared
  \item Col T: Size of the MCC that shares the maximum number of
    directors, Number of directors shared
  \item Col U: Size of the MCC that shares the minimum number of
    directors, Number of directors shared
\end{itemize}

\begin{table}[h]
  \caption{Consolidated Results for MCCs.}
  \label{tab:mcc}
  \centering
  \begin{tabular}{lllllll}
    \toprule
    \textbf{Company Name} & \textbf{P} & \textbf{Q} & \textbf{R}
      & \textbf{S} & \textbf{T} & \textbf{U} \\
    \midrule
    \textit{SS} & 720, 175    & 6, 5.0   & 7, 4  & 3, 4 & 7, 5  & 5, 2 \\
    \textit{RR} & 6863, 1955  & 5, 8.20  & 10, 3 & 5, 5 & 5, 5  & 9, 3 \\
    \textit{TT} & 912, 210    & 3, 8.67  & 16, 7 & 3, 4 & 16, 7 & 7, 1 \\
    \textit{GG} & 5044, 1527  & 5, 5.2   & 9, 6  & 3, 2 & 9, 6  & 5, 1 \\
    \textit{NN} & 21465, 6331 & 36, 6.69 & 19, 12 & 3, 3 & 10, 18 & 4, 1 \\
    \bottomrule
  \end{tabular}
\end{table}

Observations (Table~\ref{tab:mcc}):

\begin{itemize}
  \item For two MCCs with the same number of companies, the higher the
    number of shared directors, the stronger the connection between
    companies.
  \item Similarly, for two MCCs with the same number of shared
    directors, the fewer the number of companies in the MCC, the
    stronger the connection between companies.
  \item A stronger MCC can exert more influence compared to a weaker
    MCC.
\end{itemize}

\subsection{Maximal Frequent Director Itemsets (MFDIs)}

\begin{figure}[ht]
  \centering
  \includegraphics[width=0.43\textwidth]{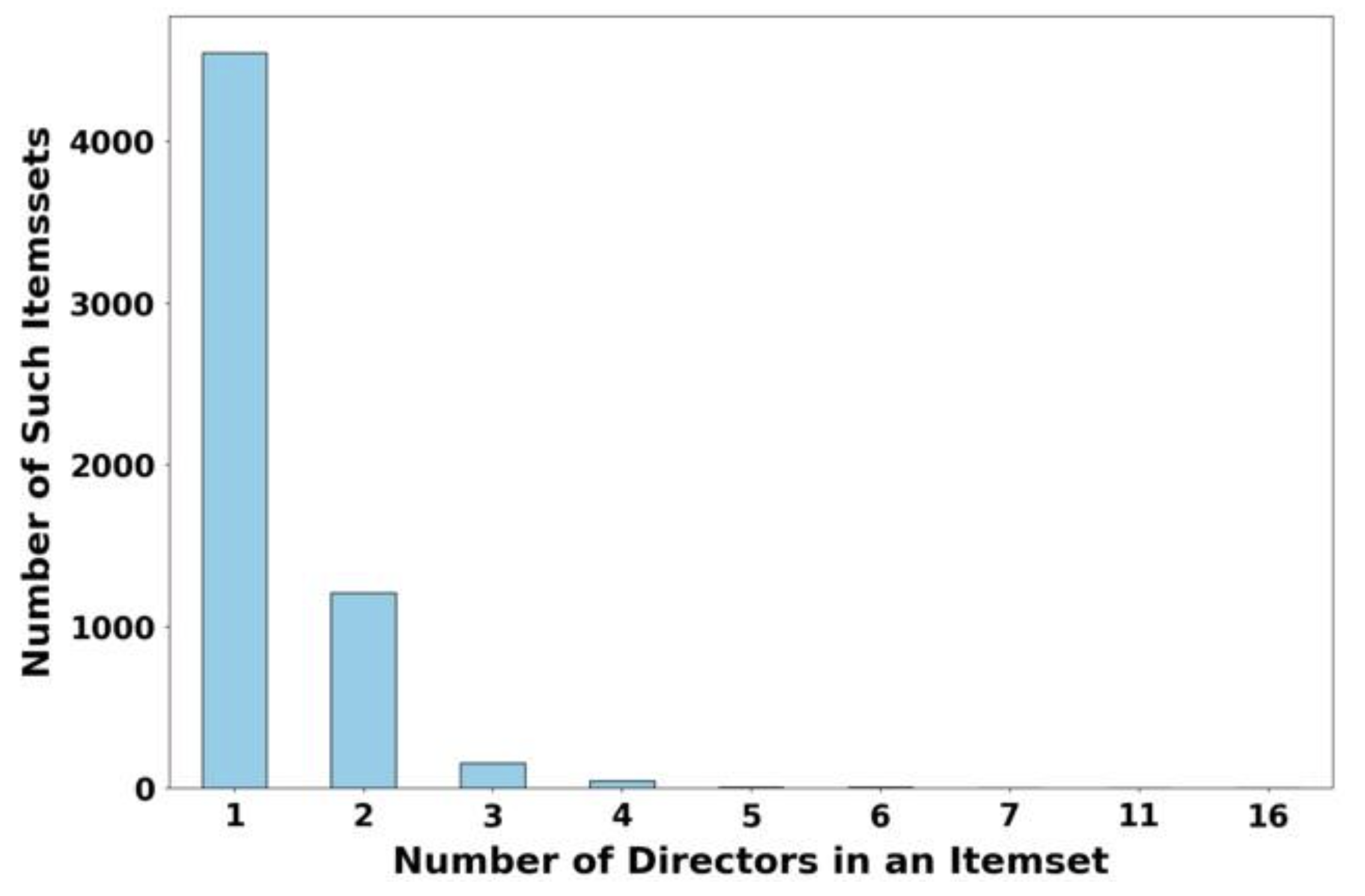}
  \includegraphics[width=0.45\textwidth]{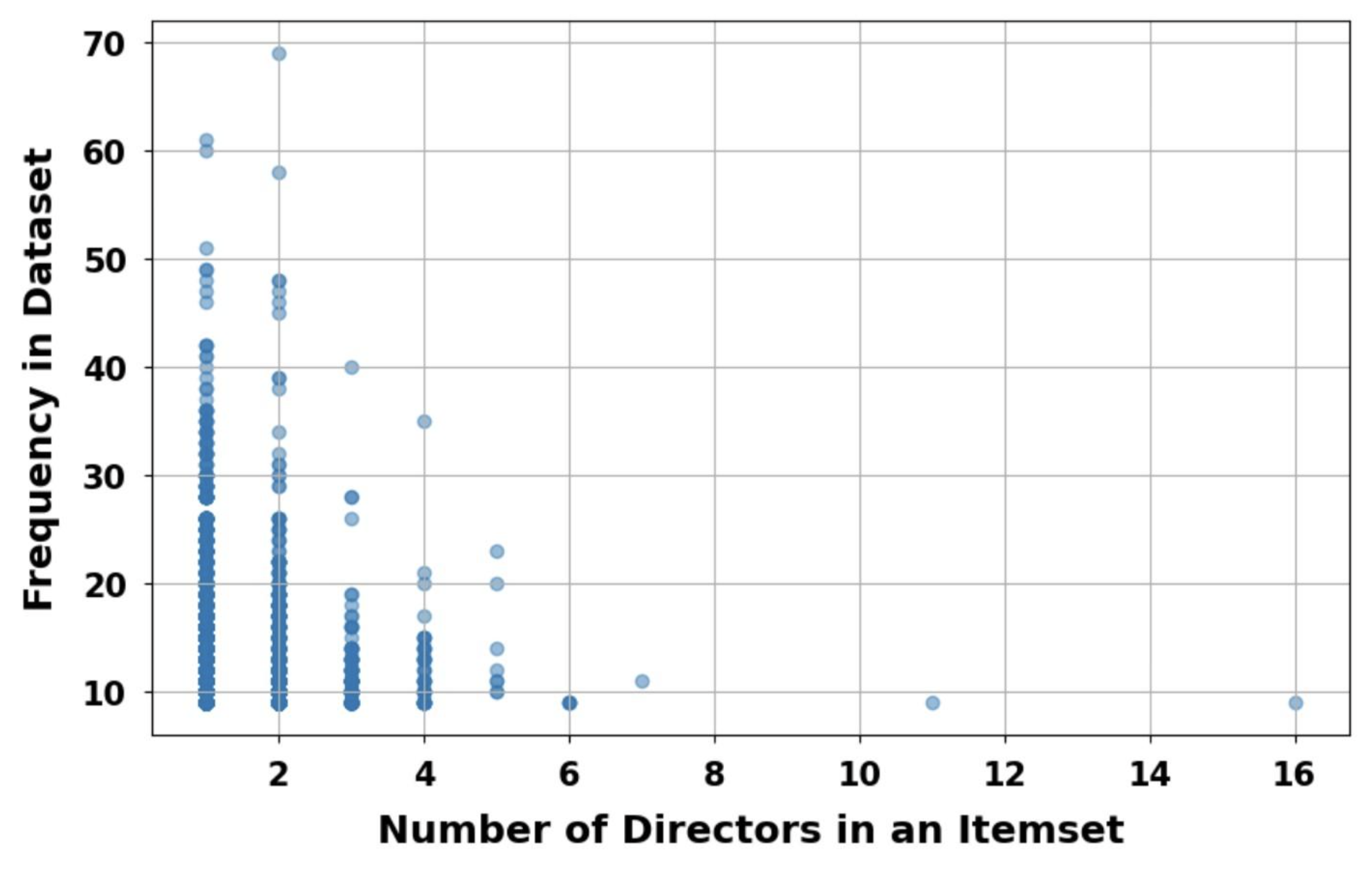}
  \caption{\centering General Distribution Plots for MFDIs. (a) Frequency of Number of Directors in an Itemset\quad (b) Number of Directors in Itemset vs Frequency of Itemset in Data.}
  \label{fig:mfdi-dist}
\end{figure}

We extracted Maximal Frequent Itemsets (MFDIs) of directors using the
pipeline outlined above and the Company Information file. The total
number of companies is 87,187, with a minimum support of 0.0001. Hence,
the minimum number of times a director itemset needs to occur to be
considered frequent is 9 ($0.0001 \times 87{,}187$). A total of 5,972
MFDIs were extracted. Figure~\ref{fig:mfdi-dist} illustrates the general
distribution of the extracted MFDIs. Figure~\ref{fig:mfdi-dist}(a) shows
a frequency graph of the number of directors in an itemset, while
Figure~\ref{fig:mfdi-dist}(b) shows a distribution graph of the number of items in an itemset and its frequency in the dataset. The frequency of an itemset here refers to the number of times the itemset occurs in the dataset, which corresponds to the number of companies shared between its corresponding directors.

Table~\ref{tab:mfdi-support} and Table~\ref{tab:mfdi-size} summarize the
top 5 results of MFDIs sorted by support and number of directors in the
itemset, respectively. In both tables, the ``Support (Freq)'' column
denotes the support value the itemset carries. The value in parentheses
is the number of times the itemset occurs in the Company Information
file, representing the number of distinct companies shared by the directors listed in the ``Frequent Director Itemset'' column.

\begin{table}[h]
  \caption{Top 5 MFDIs Based on Frequency or Support.}
  \label{tab:mfdi-support}
  \centering
  \begin{tabular}{ll}
    \toprule
    \textbf{Support (Freq)} & \textbf{Frequent Director Itemset} \\
    \midrule
    $0.000791402\ (69)$ & 2 Directors \\
    $0.000665237\ (58)$ & 2 Directors \\
    $0.000550541\ (48)$ & 2 Directors \\
    $0.000550541\ (48)$ & 2 Directors \\
    $0.000539071\ (47)$ & 2 Directors \\
    \bottomrule
  \end{tabular}
\end{table}

Observation (Table~\ref{tab:mfdi-support}):

\begin{itemize}
  \item \textit{Four} of the top \textit{five} results in
    Table~\ref{tab:mfdi-support} are family owned.
  \item \textit{Four} of the top \textit{five} results in
    Table~\ref{tab:mfdi-support} are in Real Estate/Infrastructure
    Development Business.
\end{itemize}

\begin{table}[h]
  \caption{Top 5 MFDIs Based on Number of Directors in Frequent
    Itemsets.}
  \label{tab:mfdi-size}
  \centering
  \begin{tabular}{ll}
    \toprule
    \textbf{Support (Freq)} & \textbf{Frequent Director Itemset} \\
    \midrule
    $0.000103226\ (9)$  & 16 Directors \\
    $0.000103226\ (9)$  & 11 Directors \\
    $0.000126166\ (11)$ & 7 Directors  \\
    $0.000103226\ (9)$  & 6 Directors  \\
    $0.000103226\ (9)$  & 6 Directors  \\
    \bottomrule
  \end{tabular}
\end{table}

Observations (Table~\ref{tab:mfdi-size}): In row 2, of the 11 directors
8 of them shared the same last name and in row 5 all the 6 had the same
last name. Though, in Indian context, the same family name might not
necessarily translate to relationship. LLM Driver Relation Identification
showed that most family-owned businesses with relatives typically involve
wife-husband and parent-child relationships.

\subsection{Maximal Frequent Company Itemsets (MFCIs)}

We extracted Maximal Frequent Itemsets (MFCIs) of companies using the
pipeline and the Director Information file. The total number of directors
is 54,216, with a minimum support of 0.0001. Hence, the minimum number
of times a company itemset needs to occur to be considered frequent is 6
($0.0001 \times 54{,}216$). A total of 4,742 MFCIs were extracted.

\begin{figure}[ht]
  \centering
  \includegraphics[width=0.43\textwidth]{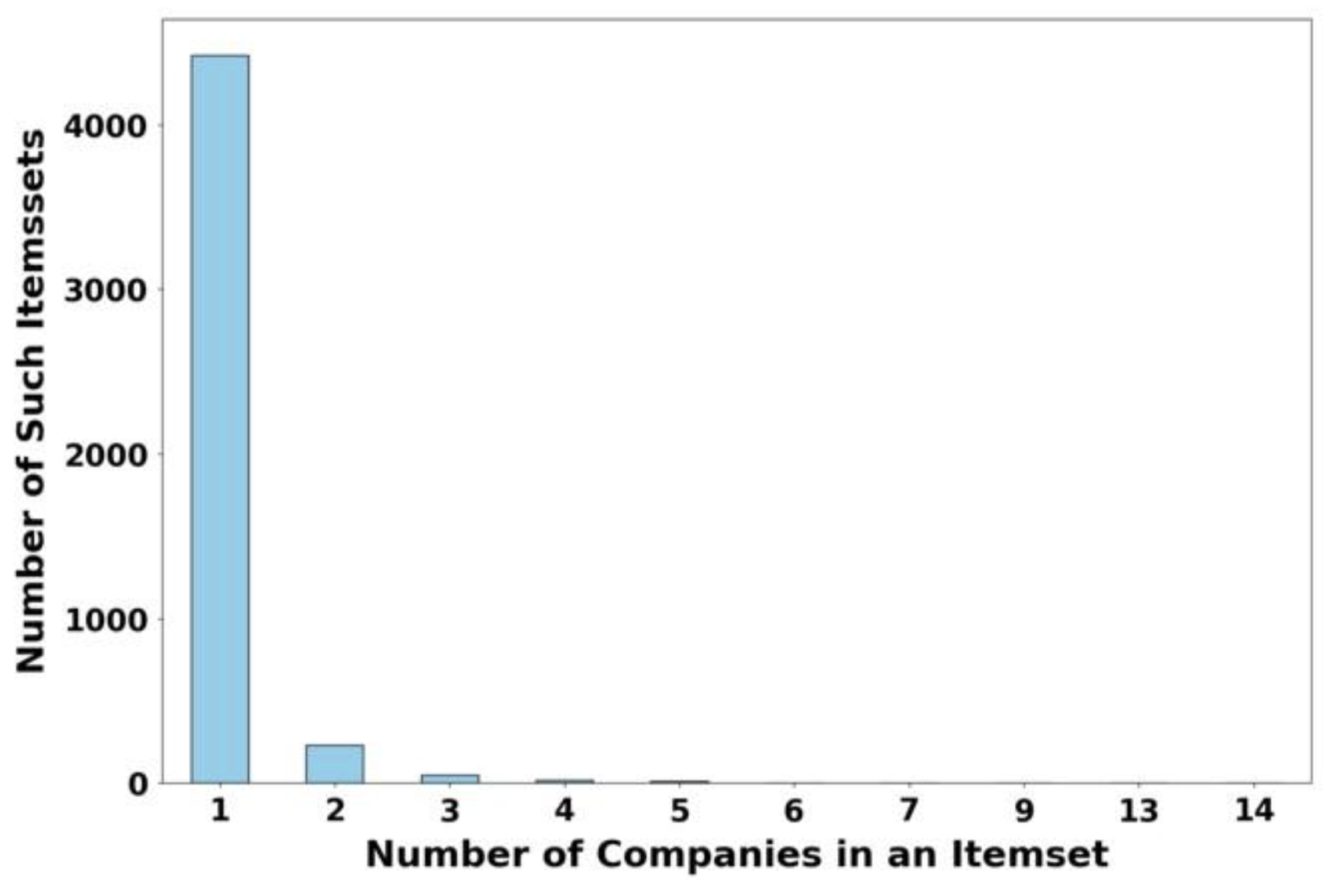}
  \includegraphics[width=0.45\textwidth]{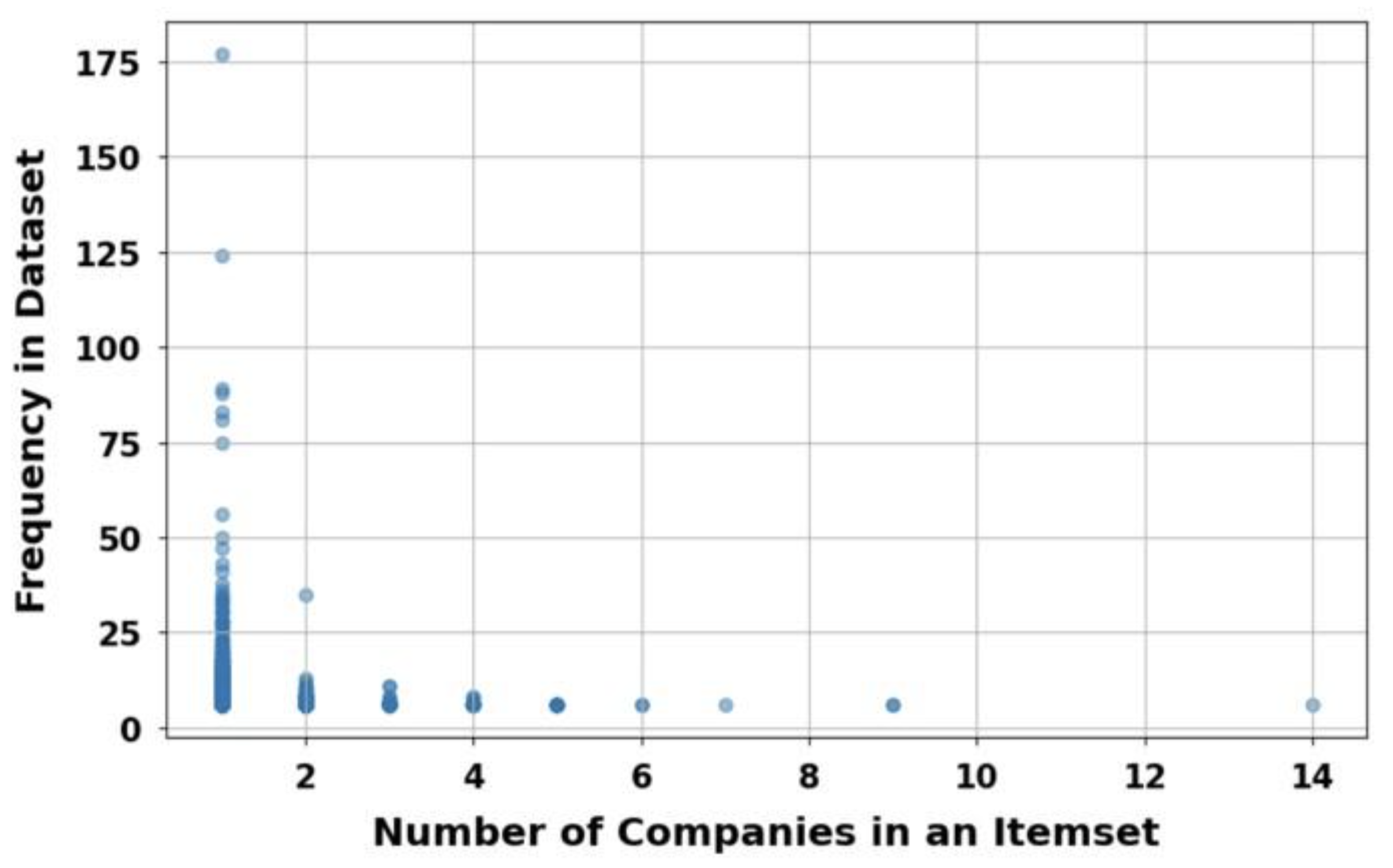}
  \caption{\centering General Distribution Plots for MFCIs. (a) Frequency of Number of Companies in an Itemset\quad (b) Number of Companies in Itemset vs Frequency of Itemset in Data.}
  \label{fig:mfci-dist}
\end{figure}

Figure~\ref{fig:mfci-dist} shows the general distribution of the
extracted MFCIs. Figure~\ref{fig:mfci-dist}(a) is a frequency graph of
the number of companies in an itemset, and Figure~\ref{fig:mfci-dist}(b)
shows a distribution graph of the number of companies in an itemset and
its frequency in the dataset. The frequency of an itemset refers to the
number of times the itemset occurs in the dataset, which corresponds to
the number of directors shared between its corresponding companies.

From the top 5 results of MFCIs sorted by support/frequency and the
number of companies in the itemset, respectively, the names of eminent
industrialists from diverse fields are directors in one company was
observed. We were also able to extract 5 MFCIs with 46 companies having
a pair of directors sharing the same last name.

The examination of corporate networks goes beyond simply identifying
connected individuals. The next step is uncovering latent connections
and underlying factors that contribute to these director-director
associations. By employing web data scraping techniques
\citep{sancheti2024}, we extract meaningful insights regarding the
interrelations between directors. These connections span a range of
relationships, from personal affiliations, such as familial bonds, to
professional ties, including shared work histories, concurrent
memberships in professional organizations, and other affiliations within
the professional sphere.

\subsection{Personal Relation Identification}

To investigate personal relations between two directors, we employ a
diverse array of web search techniques to gather information about their
personal affiliations available across search engines. By utilizing
various search queries such as ``Director 1, Director 2,'' ``Director 1,
Director 2 relation,'' or ``Director 1, Director 2 Family tree,'' we
retrieve the top five search results and extract textual information
through web scraping. To ensure exact matches with the directors' names,
we apply several search engine optimization techniques. Inspired by
prompt engineering methods suggested by \citet{wang2023} and
\citet{ashok2023}, we develop a specialized prompt for identifying
personal relations between directors within text data.

\textit{Prompt Design to Extract Personal Relation Between Directors
from Text:}

\begin{enumerate}[label=\alph*)]
  \item \textbf{Task Description:}\\
    You are a linguistic analyst. The task is to analyze a given text
    and determine if there is any familial relationship implied between
    \{Director 1\} and \{Director 2\}.

  \item \textbf{Predefined Requirements:}\\
    Use this list to return the identified relation:

    ``husband - wife''; ``daughter - father''; ``nephew - uncle'';
    ``mother - son''; ``sister - brother''; ``grandfather -
    granddaughter''; ``grandmother - grandson''; ``cousin - cousin'';
    ``aunt - nephew''; ``stepmother - stepson''; ``stepfather -
    stepdaughter''; ``godmother - godson''; ``adoptive mother - adopted
    son''; ``sister-in-law - brother-in-law''; ``friend - friend''

  \item \textbf{Edge Case Handling:}
    \begin{itemize}
      \item If the text does not mention either director, return the
        answer as `Not Available'.
      \item If the text mentions both directors but does not imply any
        familial relationship between them, return the answer as `Not
        Available'.
    \end{itemize}

  \item \textbf{Output Formatting:}\\
    Return the identified familial relationship in JSON format with the
    key as: ``Relation'': ``husband - wife''

  \item \textbf{Self-Verification and Reinforced Learning:}

    To ensure that the identified relationship is based solely on
    explicit textual evidence and not just based on the director names.
    The carefully devised prompt above ensures consistency and
    standardization, with the model specifically trained to recognize a
    predefined list of 15 relation pairs, such as Wife -- Husband,
    Daughter -- Father, Nephew -- Uncle, etc. By attaching the scraped
    textual data along with this prompt and sending the API request to
    OpenAI's GPT-3.5-turbo, we obtain the desired output, indicating
    any inference regarding familial ties within the provided text. If
    no such relationship is identified, the model will return `Not
    Available'. In such cases, we proceed to investigate professional
    links between the directors. The above run on 2 directors with no
    common last name was able to extract the relationship of
    `uncle-nephew'. The directors selected were from the same
    conglomerate ABCD, which has been used for testing the pipeline.
\end{enumerate}

\subsection{Professional Relation Identification}

To extract professional links between directors, we leverage the Web
Profile Extraction pipeline \citep{sancheti2024}. This systematic
process involves web searches, data scraping, LLM-driven Named Entity
Recognition (NER), and standardization of information sourced from
Wikipedia. Using this pipeline, we extract structured web profiles in
JSON format for both directors to analyze probable professional links.
All extracted entities are standardized using Wikipedia to address the
need for homogenizing their diverse representations. By identifying
common Wikipedia links for institution or organization names across these
profiles, we can detect matches that indicate professional affiliations.
The method was tested on 2 directors of a large service company in India
and we were able to extract that both went to same college and worked at
the same company (different periods) before founding the company.

\section{Discussion}

This study proposes a systematic pipeline for constructing a
\textit{directorship connectome}, defined as a graph-based
representation of interlocking directorates. The graph can comprise of
nodes representing directors and companies---with edges representing
directorship affiliations or inter-director relation. A Breadth-First
Search (BFS) algorithm is employed to traverse the network and extract
relevant connections. To further characterize the structural properties
of the network, clique analysis is conducted, allowing for the assessment
of edge density and the relative strength of inter-organizational or
interpersonal ties. Leveraging LLM-driven Relation Identification, we
successfully untangled complex inter-director relationships, uncovering
both familial and professional ties.

Applying it on a dataset, which spans approximately 55,000 distinct
directors across 37,123 companies (about 30\% of the total dataset), we
observed that 17\% of individuals are directors in two companies, and
58.6\% of directors hold positions at multiple companies. Of particular
relevance to Indian business houses, our analysis found that at least one
pair of directors sharing the same last name appeared in 30\% of
companies. While shared surnames do not necessarily indicate familial
ties, they could reflect community affiliation, caste-based connections,
or generational naming conventions, such as a father's first name
becoming a child's last name. These findings highlight the importance of
a thorough data analysis pipeline to understand director interlocks.
Identifying frequently interlocked directors or companies provides
valuable insights into market power and corporate structures. To the
best of our knowledge, our work is the first to present a method to
extract and prepare the interlocks for visualization and analysis. The
findings demonstrates that the proposed methodologies not only uncover
hidden interlocks but also provide insight into the structural
influences---such as familial ties, professional networks, and industry
types---that shape these connections. For instance, the pipeline was able
to extract a previously non-obvious connection---namely, a collegiate
friendship between two founders of a major service-sector company---which
would have otherwise required a time-consuming and manual web-based
investigation. Additionally, the analysis of maximal frequent itemsets
revealed an intriguing pattern: several prominent industrialists from
diverse sectors hold board positions within a single company, indicating
a strategic alignment. Analysis of first-order and second-order (direct
and indirect) company--company graphs reveals the role of a specific
company functioning as a \textit{connector} or \textit{hub} linking
multiple other firms. This structural position suggests a potential for
disproportionate influence or control over connected entities,
highlighting the company's centrality within the broader
inter-organizational network. This deeper understanding is crucial for
assessing the broader implications of director interlocks on corporate
strategies, governance, and market behaviour.

Ethical considerations are also central to this study, particularly
regarding concentrated power in corporate networks. Research from the
last decade shows that interlocking directorships, even between competing
firms, can enhance sustainability and social performance \citep{feng2024},
influence accounting policies \citep{karim2022}, reduce misconduct
\citep{wang2022}, enhance performance by access to resources
\citep{hillman2011} but may also lead to reduced R\&D investment
\citep{sierraMoran2024}. A comprehensive analysis \citep{ma2024} reveals
the impact of information transfer on corporate strategies, though it
also emphasizes the challenges of establishing causality due to
endogeneity, omitted variables, and reverse causality.
\citet{edacherian2024} examined interlocking directorates and firm
performance in Indian companies, finding that performance suffers when
directors assigned to auditing roles also hold similar responsibilities
in their parent companies. While interlocks can facilitate knowledge
transfer and sustainability, they also raise concerns about conflicts of
interest, market dominance, and potential undue governmental influence.

\section{Conclusion}

The findings from the Indian conglomerate underscore the value of
data-driven analysis in informing ethical and regulatory frameworks
governing corporate directorship. Specifically, the results point to a
need for greater clarity and accountability in the responsibilities of
directors, as well as enhanced oversight mechanisms by regulatory bodies
to ensure transparent and ethical governance practices. The preliminary
analysis on only 30\% of the total dataset opens several avenues for
future exploration with complete data, including expanding the
methodology to other organizational structures (for example: academic or
research institutes) and refining the analysis to address endogeneity
and reverse causality issues. Ultimately, this work underscores the
importance of an informed, ethical, and strategic approach to
understanding director interlocks. By uncovering the structures behind
these interlocks, we contribute to a deeper understanding of corporate
governance, market influence, and the broader power dynamics in the
business world.

Despite its insights, this study has limitations, such as potential
dataset biases and challenges in capturing the full complexity of
inter-director relationships. Addressing these gaps will be crucial for
advancing our understanding of corporate networks in diverse cultural and
economic contexts. Running the pipeline on a larger dataset recorded by
the ministry of commerce, a government agency will help extract more
director `connectomes' (interlocks).

\newpage
\bibliographystyle{unsrtnat}
\bibliography{references}

@incollection{adams2017,
  title     = {Boards, and the directors who sit on them},
  author    = {Adams, Ren{\'e}e B.},
  booktitle = {The Handbook of the Economics of Corporate Governance},
  volume    = {1},
  pages     = {291--382},
  year      = {2017},
  publisher = {North-Holland},
  doi       = {10.1016/bs.hecg.2017.11.002}
}

@misc{ashok2023,
  title         = {{PromptNER}: Prompting for Named Entity Recognition},
  author        = {Ashok, Dhananjay and Lipton, Zachary C.},
  year          = {2023},
  eprint        = {2305.15444},
  archiveprefix = {arXiv},
  primaryclass  = {cs.CL},
  url           = {https://arxiv.org/abs/2305.15444}
}

@article{barone2025,
  title   = {Interlocking Directorates and Competition in Banking},
  author  = {Barone, Guglielmo and Schivardi, Fabiano and Sette, Enrico},
  journal = {The Journal of Finance},
  volume  = {80},
  number  = {4},
  pages   = {1963--2016},
  year    = {2025},
  doi     = {10.1111/jofi.13426}
}

@article{bron1973,
  title   = {Algorithm 457: Finding All Cliques of an Undirected Graph},
  author  = {Bron, Coen and Kerbosch, Joep},
  journal = {Communications of the ACM},
  volume  = {16},
  number  = {9},
  pages   = {575--577},
  year    = {1973},
  doi     = {10.1145/361336.361338}
}

@article{caiazza2019,
  title   = {An Institutional Contingency Perspective of Interlocking
             Directorates},
  author  = {Caiazza, Rosa and Cannella Jr., Albert A. and Phan, Phillip H.
             and Simoni, Michele},
  journal = {International Journal of Management Reviews},
  volume  = {21},
  number  = {3},
  pages   = {277--293},
  year    = {2019},
  doi     = {10.1111/ijmr.12192}
}

@article{cazals2008,
  title   = {A Note on the Problem of Reporting Maximal Cliques},
  author  = {Cazals, Fr{\'e}d{\'e}ric and Karande, Chinmay},
  journal = {Theoretical Computer Science},
  volume  = {407},
  number  = {1--3},
  pages   = {564--568},
  year    = {2008},
  doi     = {10.1016/j.tcs.2008.05.010}
}

@article{dooley1969,
  title   = {The Interlocking Directorate},
  author  = {Dooley, Peter C.},
  journal = {The American Economic Review},
  volume  = {59},
  number  = {3},
  pages   = {314--323},
  year    = {1969},
  url     = {http://www.jstor.org/stable/1808960}
}

@article{edacherian2024,
  title   = {Connecting the Right Knots: The Impact of Board Committee
             Interlocks on the Performance of Indian Firms},
  author  = {Edacherian, Saju and Richter, Ansgar and Karna, Amit and
             Gopalakrishnan, Bharath},
  journal = {Corporate Governance: An International Review},
  volume  = {32},
  number  = {1},
  pages   = {135--155},
  year    = {2024},
  doi     = {10.1111/corg.12534}
}

@article{feng2024,
  title   = {Does a Company's Position Within the Interlocking Director
             Network Influence Its {ESG} Performance?---Empirical Evidence
             from Chinese Listed Companies},
  author  = {Feng, Haotian and Zhang, Zhe and Wang, Qian and Yang, Lei},
  journal = {Sustainability},
  volume  = {16},
  number  = {10},
  pages   = {4190},
  year    = {2024},
  doi     = {10.3390/su16104190}
}

@inproceedings{grahne2003,
  title     = {High Performance Mining of Maximal Frequent Itemsets},
  author    = {Grahne, G{\"o}sta and Zhu, Jianfei},
  booktitle = {6th International Workshop on High Performance Data Mining},
  volume    = {16},
  pages     = {34},
  year      = {2003}
}

@inproceedings{han2000,
  title     = {Mining Frequent Patterns without Candidate Generation},
  author    = {Han, Jiawei and Pei, Jian and Yin, Yiwen},
  booktitle = {Proceedings of the 2000 ACM SIGMOD International Conference
               on Management of Data},
  volume    = {29},

  pages     = {1--12},
  year      = {2000},
  doi       = {10.1145/335191.335372}
}

@article{helmers2017,
  title   = {Do Board Interlocks Increase Innovation? Evidence from a
             Corporate Governance Reform in India},
  author  = {Helmers, Christian and Patnam, Manasa and Rau, P. Raghavendra},
  journal = {Journal of Banking \& Finance},
  volume  = {80},
  pages   = {51--70},
  year    = {2017},
  doi     = {10.1016/j.jbankfin.2017.03.016}
}

@article{hernandez2019,
  title   = {The Impact of Interlocking Directorates on Innovation: The
             Effects of Business and Social Ties},
  author  = {Hern{\'a}ndez-Lara, Ana Beln and Gonzales-Bustos, Juan Pablo},
  journal = {Management Decision},
  volume  = {57},
  number  = {10},
  pages   = {2799--2815},
  year    = {2019},
  doi     = {10.1108/MD-10-2017-0996}
}

@article{hillman1999,
  title   = {Corporate Political Strategy Formulation: A Model of Approach,
             Participation, and Strategy Decisions},
  author  = {Hillman, Amy J. and Hitt, Michael A.},
  journal = {Academy of Management Review},
  volume  = {24},
  number  = {4},
  pages   = {825--842},
  year    = {1999},
  doi     = {10.5465/amr.1999.2553256}
}

@article{hillman2011,
  title   = {What {I} Like About You: A Multilevel Study of Shareholder
             Discontent with Director Monitoring},
  author  = {Hillman, Amy J. and Shropshire, Christine and Certo, S. Trevis
             and Dalton, Dan R. and Dalton, Catherine M.},
  journal = {Organization Science},
  volume  = {22},
  number  = {3},
  pages   = {675--687},
  year    = {2011},
  doi     = {10.1287/orsc.1100.0542}
}

@article{holburn2008,
  title   = {Making Friends in Hostile Environments: Political Strategy in
             Regulated Industries},
  author  = {Holburn, Guy L. F. and Vanden Bergh, Richard G.},
  journal = {Academy of Management Review},
  volume  = {33},
  number  = {2},
  pages   = {521--540},
  year    = {2008},
  doi     = {10.5465/amr.2008.31193530}
}

@inproceedings{hua2020,
  title     = {List All Maximal Cliques of an Undirected Graph: A Parallable
               Algorithm},
  author    = {Hua, Xiao and Zhong, Ming and Liu, Quan and Wang, Meng},
  booktitle = {{IOP} Conference Series: Materials Science and Engineering},
  volume    = {790},

  pages     = {012076},
  year      = {2020},
  publisher = {IOP Publishing},
  doi       = {10.1088/1757-899X/790/1/012076}
}

@article{jackling2009,
  title   = {Board Structure and Firm Performance: Evidence from {India}'s
             Top Companies},
  author  = {Jackling, Beverley and Johl, Shireenjit},
  journal = {Corporate Governance: An International Review},
  volume  = {17},
  number  = {4},
  pages   = {492--509},
  year    = {2009},
  doi     = {10.1111/j.1467-8683.2009.00760.x}
}

@article{karim2022,
  title   = {Do Directors Have Style? Board Interlock and Accounting
             Properties},
  author  = {Karim, Khondkar E. and Li, Jing and Lin, Karen Jingrong and
             Robin, Ashok},
  journal = {Journal of Business Finance \& Accounting},
  volume  = {49},
  number  = {1--2},
  pages   = {3--32},
  year    = {2022},
  doi     = {10.1111/jbfa.12567}
}

@article{lamb2016,
  title   = {The ``Ties That Bind'' Board Interlocks Research: A Systematic
             Review},
  author  = {Lamb, Nicholas H. and Roundy, Philip},
  journal = {Management Research Review},
  volume  = {39},
  number  = {11},
  pages   = {1516--1542},
  year    = {2016},
  doi     = {10.1108/MRR-08-2015-0188}
}

@article{loderer2002,
  title   = {Board Overlap, Seat Accumulation and Share Prices},
  author  = {Loderer, Claudio and Peyer, Urs},
  journal = {European Financial Management},
  volume  = {8},
  number  = {2},
  pages   = {165--192},
  year    = {2002},
  doi     = {10.1111/1468-036X.00183}
}

@article{ma2024,
  title   = {Director Interlocks: Information Transfer in Board Networks},
  author  = {Ma, Zhuang and Shi, Lin and Yu, Kun and Zhou, Nan},
  journal = {Encyclopedia},
  volume  = {4},
  number  = {1},
  pages   = {117--124},
  year    = {2024},
  doi     = {10.3390/encyclopedia4010009}
}

@article{mazzola2016,
  title   = {The Interaction Between Inter-firm and Interlocking Directorate
             Networks on Firm's New Product Development Outcomes},
  author  = {Mazzola, Erica and Perrone, Giovanni and Kamuriwo, Dzidzai
             Samuel},
  journal = {Journal of Business Research},
  volume  = {69},
  number  = {2},
  pages   = {672--682},
  year    = {2016},
  doi     = {10.1016/j.jbusres.2015.08.033}
}

@article{mizruchi1996,
  title   = {What Do Interlocks Do? An Analysis, Critique, and Assessment
             of Research on Interlocking Directorates},
  author  = {Mizruchi, Mark S.},
  journal = {Annual Review of Sociology},
  volume  = {22},
  number  = {1},
  pages   = {271--298},
  year    = {1996},
  doi     = {10.1146/annurev.soc.22.1.271}
}

@article{moon1965,
  title   = {On Cliques in Graphs},
  author  = {Moon, John W. and Moser, Leo},
  journal = {Israel Journal of Mathematics},
  volume  = {3},
  pages   = {23--28},
  year    = {1965},
  doi     = {10.1007/BF02760024}
}

@article{ozmel2013,
  title   = {Signals Across Multiple Networks: How Venture Capital and
             Alliance Networks Affect Interorganizational Collaboration},
  author  = {Ozmel, Umit and Reuer, Jeffrey J. and Gulati, Ranjay},
  journal = {Academy of Management Journal},
  volume  = {56},
  number  = {3},
  pages   = {852--866},
  year    = {2013},
  doi     = {10.5465/amj.2010.0838}
}

@article{palmer1986,
  title   = {The Ties That Bind: Organizational and Class Bases of
             Stability in a Corporate Interlock Network},
  author  = {Palmer, Donald and Friedland, Roger and Singh, Jitendra V.},
  journal = {American Sociological Review},
  pages   = {781--796},
  year    = {1986},
  doi     = {10.2307/2095369}
}

@article{pfeffer1972,
  title   = {Size and Composition of Corporate Boards of Directors: The
             Organization and Its Environment},
  author  = {Pfeffer, Jeffrey},
  journal = {Administrative Science Quarterly},
  volume  = {17},
  number  = {2},
  pages   = {218--228},
  year    = {1972},
  doi     = {10.2307/2393956}
}

@book{pfeffer1978,
  title     = {The External Control of Organizations: A Resource Dependence
               Perspective},
  author    = {Pfeffer, Jeffrey and Salancik, Gerald R.},
  year      = {1978},
  publisher = {Harper \& Row},
  address   = {New York}
}

@inproceedings{sancheti2024,
  title     = {{LLM} Driven Web Profile Extraction for Identical Names},
  author    = {Sancheti, Pranav and Karlapalem, Kamalakar and Vemuri,
               Kavitha},
  booktitle = {Companion Proceedings of the ACM Web Conference 2024
               ({WWW} '24)},
  year      = {2024},
  doi       = {10.1145/3589335.3651573}
}

@incollection{sapinski2018,
  title     = {Interlocking Directorates and Corporate Networks},
  author    = {Sapinski, Jean-Philippe and Carroll, William K.},
  booktitle = {Handbook of the International Political Economy of the
               Corporation},
  pages     = {45--60},
  year      = {2018},
  publisher = {Edward Elgar Publishing}
}

@incollection{paoloni2025,
  title     = {Interlocking Directorates and Gender Inclusion: Unveiling
               the Role of Women in the Italian Listed Companies},
  author    = {Santolamazza, Valentina and Paoloni, Nicola and Elia,
               Brunella and Paoloni, Massimo},
  booktitle = {Shaping Tomorrow: Gender Perspectives in a Sustainable
               World},
  pages     = {279--293},
  year      = {2025},
  publisher = {Springer Nature Switzerland},
  address   = {Cham},
  doi       = {10.1007/978-3-031-84226-5_18}
}

@article{sierraMoran2024,
  title   = {The Moderating Effect of Interlocking Directors on the
             Relationship Between {R\&D} Investments and Firm Value},
  author  = {Sierra-Mor{\'a}n, Javier and Cabeza-Garc{\'i}a, Laura and
             Gonz{\'a}lez-{\'A}lvarez, Nuria},
  journal = {The Journal of Technology Transfer},
  volume  = {49},
  number  = {3},
  pages   = {1016--1042},
  year    = {2024},
  doi     = {10.1007/s10961-023-10021-z}
}

@article{singh2025,
  title   = {Effect of Director Interlocks, Composition of Board and
             Family Involvement on {IPO} Underpricing: An Empirical Study
             of {IPOs} in India},
  author  = {Singh, Abhishek Kumar and Rani, Poonam and Gour, Rohit},
  journal = {Management and Labour Studies},
  pages   = {0258042X251336527},
  year    = {2025},
  doi     = {10.1177/0258042X251336527}
}

@article{stearns1986,
  title   = {Broken-Tie Reconstitution and the Functions of
             Interorganizational Interlocks: A Reexamination},
  author  = {Stearns, Linda Brewster and Mizruchi, Mark S.},
  journal = {Administrative Science Quarterly},
  pages   = {522--538},
  year    = {1986},
  doi     = {10.2307/2392924}
}

@inproceedings{tomita2004,
  title     = {The Worst-Case Time Complexity for Generating All Maximal
               Cliques},
  author    = {Tomita, Etsuji and Tanaka, Akira and Takahashi, Haruhisa},
  booktitle = {International Computing and Combinatorics Conference},
  pages     = {161--170},
  year      = {2004},
  publisher = {Springer Berlin Heidelberg},
  address   = {Berlin, Heidelberg},
  doi       = {10.1007/978-3-540-27798-9_19}
}

@misc{wang2023,
  title         = {{GPT-NER}: Named Entity Recognition via Large Language
                  Models},
  author        = {Wang, Shuhe and Sun, Xiaofei and Li, Xiaoya and Ouyang,
                   Rongbin and Wu, Fei and Zhang, Tianwei and Li, Jiwei and
                   Wang, Guoyin},
  year          = {2023},
  eprint        = {2304.10428},
  archiveprefix = {arXiv},
  primaryclass  = {cs.CL},
  url           = {https://arxiv.org/abs/2304.10428}
}

@article{wang2022,
  title   = {Learning from Failures: Director Interlocks and Corporate
             Misconduct},
  author  = {Wang, Zijun and Yao, Shouyu and Sensoy, Ahmet and Goodell,
             John W. and Cheng, Feiyang},
  journal = {International Review of Financial Analysis},
  volume  = {84},
  pages   = {102406},
  year    = {2022},
  doi     = {10.1016/j.irfa.2022.102406}
}

@article{zhang2025,
  title   = {The Role of Interlocking Directorates and Managerial
             Characteristics on Corporate Green Innovation},
  author  = {Zhang, Xiaolan and Sun, Wei},
  journal = {Finance Research Letters},
  volume  = {74},
  pages   = {106818},
  year    = {2025},
  doi     = {10.1016/j.frl.2025.106818}
}

\end{document}